# 2D Metal Selenide-Silicon Steep Sub-Threshold Heterojunction Triodes with High On-Current Density

Jinshui Miao,[1,] Chloe Leblanc,[1] Xiwen Liu,[1] Baokun Song,[1] Huairuo Zhang,[2, 3] Sergiy Krylyuk,[2] Albert V.Davydov,[2] Tyson Back,[4] Nicholas Glavin,[4] Deep Jariwala[1,]*

[1] Department of Electrical and Systems Engineering, University of Pennsylvania, Philadelphia PA 19104, United States

[2] Materials Science and Engineering Division, National Institute of Standards and Technology, Gaithersburg, MD 20899, United States

[3] Theiss Research, Inc., La Jolla, CA 92037, United States

[4] Air Force Research Laboratory, Materials and Manufacturing Directorate, Wright-Patterson AFB, OH 45433

* Corresponding author: dmj@seas.upenn.edu

**Low power consumption in both static and dynamic modes of operation is a key requirement in modern, highly scaled nanoelectronics[1,2]. Tunneling field-effect transistors (TFETs) that exploit direct band-to-band tunneling of charges and exhibit steep sub-threshold slope (SS) transfer characteristics are an attractive option in this regard. However, current generation of Si and III-V heterojunction based TFETs while suffer from low ON current density and ON/OFF current ratios for < 60 mV/dec operation[2,3]. Semiconducting two-dimensional (2D) layers have recently renewed enthusiasm in novel device design for TFETs not only because of their atomically-thin bodies that favor superior electrostatic control[4-9] but the same feature also favors higher ON current density and consequently high ON/OFF ratio. Here, we demonstrate gate-tunable heterojunction diodes (triodes)**

**fabricated from InSe/Si 2D/3D van der Waals heterostructures, with a minimum subthreshold swing (SS) as low as 6.4 mV/dec and an SS average of 30 mV/dec over 4 decades of current. Further, the devices show a large current on/off ratio of approximately $10^6$ and on-state current density of 0.3 μA/μm at a drain bias of -1V. Our work opens new avenues for 2D semiconductors for 3D hetero-integration with Si to achieve ultra-low power logic devices.**

Dimensional scaling of metal-oxide semiconductor field-effect transistors (MOSFETs) has continued over the past decade despite diminishing returns in terms of improving packing density. A side effect of improving packing density has been power dissipation which comprises the central problem is modern, highly scaled nanoelectronics[1,2]. However, a fundamental limit on subthreshold swing (SS) in thermionic devices involving single band transport such as MOSFETs not only restricts further scaling in power consumption and supply voltage but also increases power density and dissipation in MOSFET-based circuits. This limitation on SS is set at $m*\ln(10)k_bT/q$ (~60 mV/decade at room temperature for an ideal MOSFET)[10], where $k_b$ is Boltzmann constant, m is the ideality factor (=1 for an ideal transistor), q is the elementary charge and T is temperature, arises from thermal nature of carrier injection at the metal semiconductor contact that puts a lower limit on the power consumed per switching cycle. To overcome this limitation tunneling field-effect transistors (TFETs), whose operation relies on band-to-band tunneling rather than thermionic emission, have appeared as a promising alternative to modern MOSFETs for low-power electronics.[3,4] However, TFETs have long been limited by low ON/OFF ratios for < 60 mV/dec operation and net ON current density.[2,3]

2D semiconductors have recently renewed opportunity in TFET device design because of their atomically-thin nature that permits strong electrostatic control.[4-9] This electrostatic control not only allows smaller SS values in TFETs but may also permit higher ON/OFF ratios by maximizing ON current. In addition, 2D materials are naturally self-passivated, which allows them to be easily embedded between metal gates and thin dielectrics for enabling strong electrostatic modulation. But stable and complementary doping in 2D materials remains a persistent challenge,[11,12] unlike in 3D bulk materials which have well-established, complementary doping schemes. TFETs device structures normally comprise of p-i-n homo or heterojunctions where the i (intrinsic) layer undergoes strong electrostatic modulation and shrinkage in lateral dimensions to permit direct band-to-band tunneling. In the case of intrinsic 2D materials this electrostatic modulation is relatively easy to achieve. Therefore, combining 2D materials with 3D bulk semiconductors offers an interesting avenue[13-15] for not only exploring novel TFET architectures but also exploring other advantages such a large current modulation via electrostatic gating. Here, we demonstrate unintentionally n-doped 2D-InSe and heavily p doped (p++) 3D-silicon-based gate-tunable heterojunction diodes (triodes) that overcome the thermionic limitation of conventional MOSFETs and achieve a minimum SS of 6.4 mV/decade as well as an average SS of 34 mV/decade over four orders of magnitude of drain current, simultaneously with an ON current density of 0.3 uA/um.

**Results and Discussion**

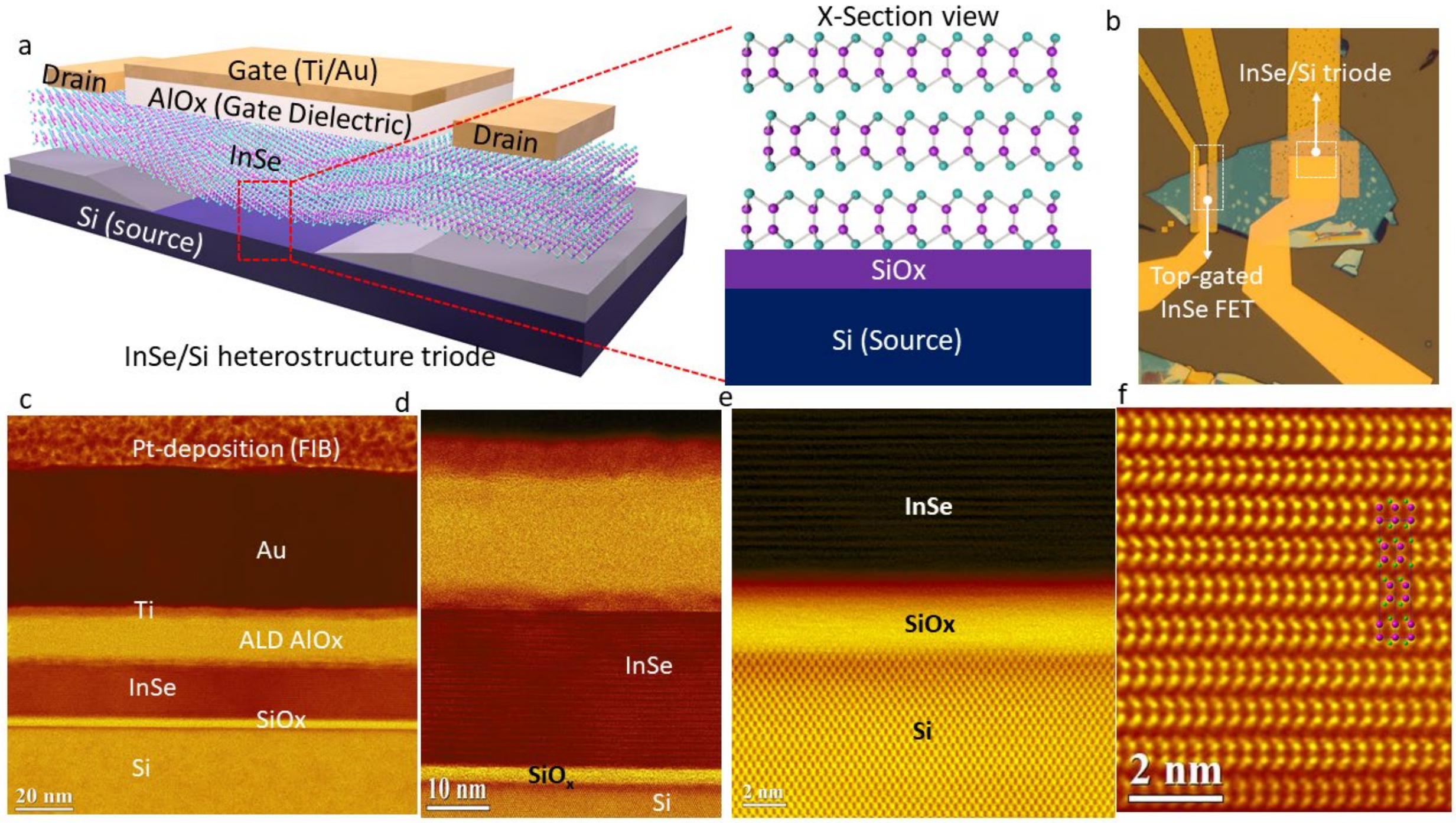

**Figure 1. 2D/3D heterojunction triode device structure and characterizations**. (a) Schematic layout of an InSe/$p^{++}$Si 2D/3D heterojunction TFET. (b) Optical micrograph of a representative 2D/3D heterojunction TFET showing InSe/Si 2D/3D heterostructure, $AlO_x$ gate dielectric, and metallic electrodes. Scale bar, 5 μm. (c) Bright-field STEM image showing the cross-sectional architecture of the device. (d-e) Zoom-in images showing the layered crystalline structure of InSe and a 2 nm-thick amorphous native silicon oxide (SiOx) between InSe and single crystalline Si. (f) Atomic resolution HAADF-STEM image overlapped with projected atomic model showing good match with the γ-InSe polytype.

We begin by etching square windows on the surface of an $SiO_2$/p-Si substrate to expose the lower layer. We then transfer mechanically exfoliated 2D InSe flakes of a few-layer-thickness onto the exposed surface (Figure 1a). The gamma phase crystalline structure is shown in **Figure 1a (X-section view) and f**. Microfabrication processes, detailed in the Methods section, are used to make the metal electrodes and gate dielectric. An optical micrograph of the InSe/p++Si 2D/3D heterojunction triodes along with control top-gated FETs is shown in Figure 1b. The cross-sectional architecture of the device is characterized by scanning transmission electron microscopy (STEM) (Fig. 1c-e). A 2 nm-thick native oxide ($SiO_x$) which was formed during the InSe transfer process is revealed between the 2D InSe and 3D Si (Fig. 1d and Fig. S1-2. The polytype structure of InSe layer is identified as γ phase with selected area electron diffraction (Fig. S1-3) and atomic resolution high angle annular dark field (HAADF) STEM analyses (Fig. 1f).

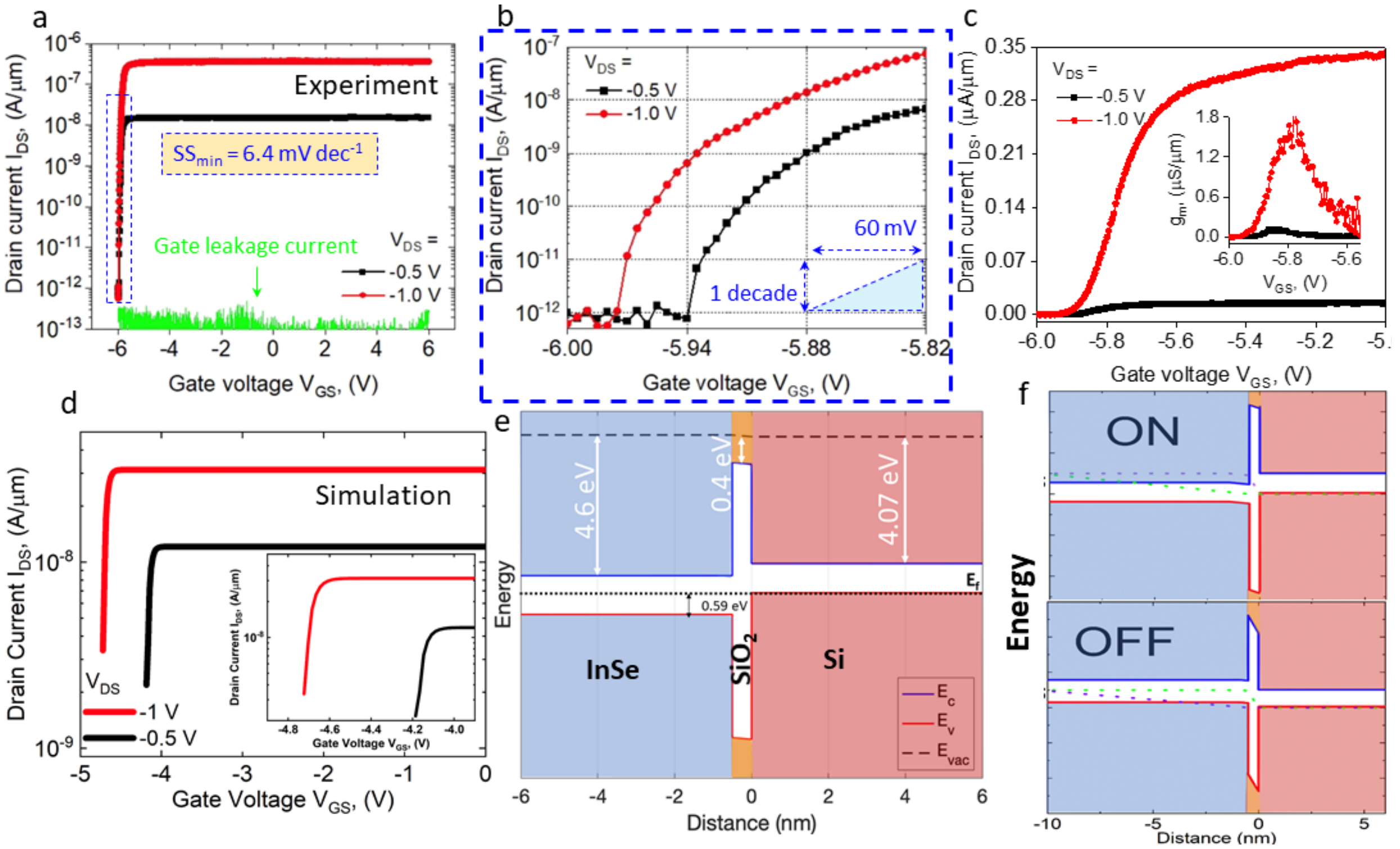

**Figure 2**. **Room-temperature electrical characteristics and band diagrams of InSe/Si heterojunction triodes.** (a) Logarithmic-scale $I_{DS}$-$V_{GS}$ transfer characteristics of InSe/Si heterojunction triode devices measured at room- temperature with applied $V_{DS}$ = -0.5 V (black) and -1.0 V (red) plots. The minimum sub-threshold swing (SS) value is 6.4 mV/decade. The green line represents gate leakage current of ~$10^{-13}$ A/μm. (b) Zoom in $I_{DS}$-$V_{GS}$ transfer characteristics (blue dashed box in panel a) showing $V_{GS}$ sweep from -6.0 to -5.8 V. The hypotenuse of right triangle is a reference for indicating the thermionic limit of 60 mV/decade for the SS. (c) Linear-scale $I_{DS}$-$V_{GS}$ transfer characteristics of the same InSe/Si heterojunction device. The inset indicates the transconductance of $I_{DS}$-$V_{GS}$ transfer curves. The transconductance shows a maximum at

-5.8 V. (d) Predicted transfer characteristics of InSe/Si heterojunction triode devices using TCAD simulator. Inset shows zoom in predicted transfer characteristics using Sentaurus TCAD simulator. (e) Simulated equilibrium band diagram at zero drain and gate bias with the three materials at the junction indicated by blue (InSe), orange ($SiO_2$) and red (Si) colors respectively while the red and blue lines indicate maxima and minima of conduction bands respectively. (f) Simulated band diagrams of ON (top) and OFF (bottom) states with the purple and green dashed lines representing quasi fermi levels for electrons and holes respectively.

We perform electrical characterization of the InSe/Si heterojunction triode in a three-terminal transistor configuration at room temperatures. The device exhibits a small SS value of 6.4 mV/decade with a current on-off ratio of $10^6$, as well as a large on-state current density of 0.3 μA/μm at an applied $V_{DS}$ = -1 V (Figure 2a). As compared to the drain current the gate-leakage current density shows negligible magnitude at ~$10^{-13}$ A/um. Figure 2b represents the zoom in transfer characteristics with $V_{GS}$ changing from -6 to -5.82 V showing the steepness in context of a magnified x- axis. It is worth noting that a gate voltage change of mere 180 mV modulates the output current by over five orders of magnitude. The same device shows a peak transconductance of 1.8 μS/μm for $V_{DS}$ = -1 V, as displayed in Figure 2c along with the linear-scale $I_{DS}$-$V_{GS}$ transfer characteristics. We note that this phenomenon is not observed in a one-off device and has been reproducible in other devices (Supporting information Figure S5). To understand the details of the device electronic structure, we have performed photoemission and Hall measurements (Supporting information Section 1) on our InSe crystals used to determine the carrier density and band alignment. Using these values combined with the known gap value of gamma phase InSe, we simulate band diagrams of near intrinsic to unintentionally n-doped InSe/p++ Si heterojunction triodes with an intermediate $SiO_2$ layer. We note that the exact structure, composition, and nature of this $SiO_2$ layer is unclear and hence modelled purely as a perfect insulator in these simulations with no traps or other leakage mechanisms present. Additional details on the composition of this $SiO_2$ interlayer are provided in Supporting information Figure S1-2. Our simulated $I_{DS}$-$V_{GS}$ characteristics (Figure 2d) show qualitative resemblance to the experimental plots as shown in Figure 2 a-b. Although the threshold and ON/OFF ratio do not match quantitatively, they are likely to vary since no trap charges at either interface have been considered in the simulation (See methods and supporting information Section S2 for simulation details). None the less the SS (< 40 mV/dec) from simulated plots (see supporting information S2), are reasonably matched to our experimentally made devices. Equilibrium band diagrams with known electron affinities and carrier concentrations shows that the InSe/$SiO_2$/Si form a type II junction with 0.59 eV valence band offset that is closely matched with photoemission results (See supporting information Figure S1-1) Simulated band diagrams of the device are also obtained in the ON and OFF states by varying the carrier density in InSe representing modulated gate bias (Figure 2 f top and bottom respectively) at a fixed drain bias of -1 V. As InSe is doped with a gate voltage the conduction band of InSe overlaps with the heavily populated valence band and doping impurity band of Si resulting is direct tunneling through the $SiO_2$ insulator. To further evaluate the electrical characteristics of our devices we make comparison with control top gated InSe MOSFETs made on the same InSe crystal as the junction as shown in Figure 1b. We observe clear and stark differences between the two devices made from the same flake (crystal) further suggesting that our heterojunction device indeed has a different charge transport mechanism i.e., BTBT. There are three clear differences observed between the control InSe FET and the InSe/Si heterojunction triode: 1. The current density of InSe MOSFET is higher than the InSe/Si heterojunction FET. 2. The average SS value in transfer characteristics of InSe FETs is more than three times lower for same oxide thickness. (Figure 3 a-b) 3. The InSe FETs have linear output characteristics while the InSe/Si heterojunctions clearly show rectifying output with gate-tunable rectification (Figure 3 c-d). First the transfer characteristics when compared over the same scale (Figure 3a) show that the InSe MOSFET (green) is clearly less steep in the sub-threshold region as compared to the InSe/Si heterojunction (blue). Further, it is also clear from the same plot that the InSe MOSFET has higher current density in the ON state as opposed to InSe/Si heterojunction triode. Interestingly both devices show very flat ON state characteristics in the semi-log scale which further suggests that the InSe/Si heterojunction can provide large swings in current with small swings in voltage and suitable for low-voltage operation. Further, taking an average slope of these $I_{DS}$-$V_{GS}$ transfer characteristics and plotting against the drain current density we find that the InSe MOSFET has SS value that hits a minimum at ~ 100 mV/dec and rapidly rises with both increase of decrease in current (See more details in Supporting information section S4). In contrast, the InSe/Si heterojunction FET shows consistently small SS value ranging from 6.4 to 60 mV/dec over 4 orders of magnitude change in current with an average of ~30 mV/dec. These average SS values have been derived using the data points of drain current in the $10^{-12}$ to $10^{-8}$ A/μm range. This is because the InSe used in our devices is unintentionally doped and near intrinsic in nature and therefore likely to show p-type conduction upon further increase in $V_{GS}$ in the negative direction (Figure 3a, green curve from -4 to -6 V). We also note that similar behavior has been obtained in intrinsic $WSe_2$/p++ Si heterojunction devices (see supporting information Section S3). Therefore, although the device is qualitatively similar to ref.[16] there are important differences between these prior works and present report. For instance, BTBT requires not only clean interfaces but "electronically" clean band-edges. For sulfide systems this is difficult to achieve and hence $MoS_2$ devices are n-doped in most cases. This is notably reduced in selenide systems which are easier to grow with near intrinsic or unintentional doping. To prove this point further, we have also made control MOSFET and InSe/Si heterojunction triode devices with substitutionally p-doped and n-doped InSe crystals (See supplementary information table 1). Neither the MOSFETs nor the heterojunctions based on intentionally doped InSe show such steep SS response as a function of gate voltage which suggests that purity/cleanliness of the crystal near band edges is of paramount importance in observing the reported steep SS behavior.

Another notable difference between the InSe MOSFETs and InSe/Si heterojunction triodes is in the output ($I_{DS}$-$V_{GS}$) characteristics. The InSe MOSFETs with Ohmic contacts show highly symmetric I-V characteristics above the origin with early signs of saturation (Figure 3c). These characteristics are representative of n-MOSFET with increasing conductance (slope) and hence current as a function of gate voltage from -6 V to 6 V in agreement with the transfer characteristics in Figure 3a. Even when magnified to small values of voltage (Figure 3c inset) these I-V curves remain linear suggesting ohmic nature of the contacts. In contrast, the InSe/Si heterojunction triode clearly demonstrates rectifying IV characteristics (Figure 3d). Not only are the characteristics rectifying but the turn on voltage of the diode is tunable as a function of gate voltage. This observation is like that of several prior reports of gate-tunable p-n diodes based on 2D materials[16-18]. When observed on a logarithmic current scale (Figure 3d, inset), the shift in turn on voltage and rise in reverse current is clear. We have further verified these output characteristics of the heterojunction as a function of doping levels in InSe via TCAD simulations and they are in good qualitative agreement with our experimental observations (see supporting information Section S3)

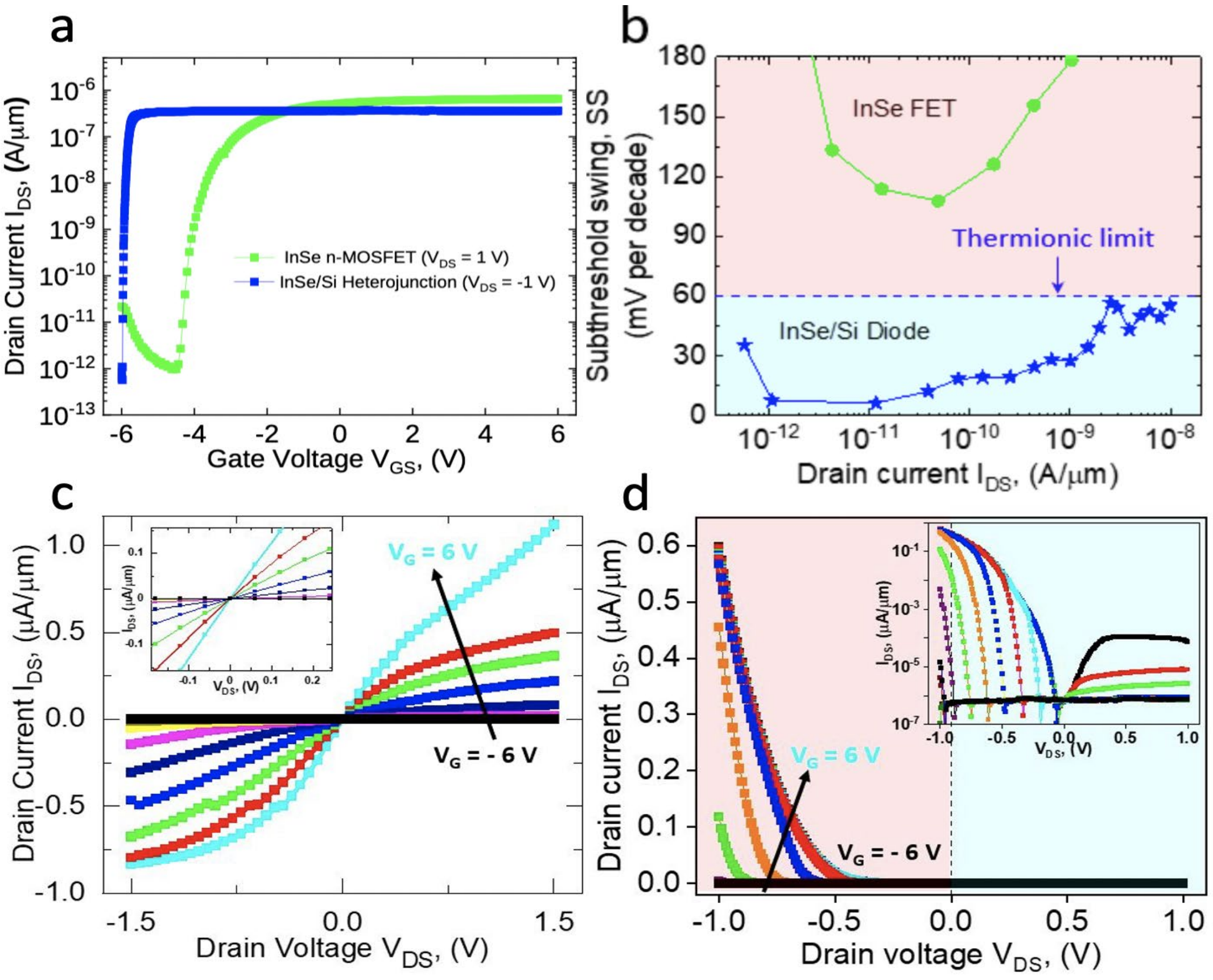

**Figure 3.** (a) Logarithmic-scale $I_{DS}$-$V_{GS}$ transfer characteristics comparison between the InSe FETs measured at $V_{DS}$ = 1V and heterojunctions at $V_{DS}$ = -1V, both at room temperature. (b) Subthreshold swing comparison of InSe/Si heterojunction FETs with control InSe FETs. The minimum SS value of control InSe FETs is around 90 mV/decade. The InSe/Si heterojunction FETs show sub-60 mV/decade over 4-5 decades of drain current. The blue dashed line represents the thermionic limit of 60 mV/decade. (c) Linear-scale $I_{DS}$-$V_{DS}$ output characteristics of control InSe FETs measured at room temperature with $V_{GS}$ varying from -6 to 6V in 1V steps. Inset shows a zoom in on the same plot closer to origin to show linearity of the characteristics. (d) Linear-scale $I_{DS}$-$V_{DS}$ output characteristics of InSe/Si heterojunctions measured at room temperature with $V_{GS}$ sweeping from -6 to 6V with 1 V step. The inset is the same plot on a logarithmic current-axis.

To further prove the differences between charge transport mechanism of the InSe MOSFETs and the InSe/Si heterojunction triodes, we perform temperature-dependent electrical characterization of the two devices as shown in Figure 4. Figure 4a shows the $I_{DS}$-$V_{GS}$ characteristics of the heterojunction device at various temperatures for $V_{DS}$ = -2 V. The threshold voltage for band-to-band tunneling (BTBT) appears to clearly shift right with reducing temperature. In addition, the magnitude of ON current plateau also appears to reduce with reducing temperatures. This suggests that the transport in these devices is complex and varies as a function of gate voltage/position of Fermi level in the InSe. Decrease in ON current as a function of T suggests some thermal barrier in the device. This thermal barrier could arise from multiple reasons ranging from imperfect metal InSe contacts to hopping transport in the non-junction part of the InSe[19,20]. Another possible mechanism to explain this is trap-assisted tunneling, where electrons first tunnel into a trap within the band gap close to the conduction band, from which they are thermally excited into the conduction band. The Bridgman-grown InSe crystals[21,22] possess interstitial atoms, vacancies and unintentional impurities. To understand this temperature activated transport in more detail we present Arrhenius plot analysis (Figure 4b) which shows plots made for different gate voltages. We observe an activation energy of 0.2 eV extracted for the subthreshold region. Further this activation energy changes from 0.2 to 0.05 eV as the gate voltage changes from -5.36 to -4.0 V suggesting that the transport is weakly dependent on temperature in the high doping density ($V_{GS}$ = -4 V). Given the steep drop in current and lack of ability to obtain currents below $10^{-11}$ A it is difficult to ascertain the nature of temperature dependence in the sub-threshold region. However, it is well known for thermionic transport in conventional MOSFETs that the SS depends linearly on T as $SS = m \cdot \ln(10) k_b T/q$, where m is the ideality factor (~1 for an ideal MOSFET). Upon plotting the SS slope as function of T we find that the slope is nearly independent of T in stark contrast with InSe MOSFET which shows as clear dependence on T with an m factor > 2 (Figure 4 c). This provides strong evidence that the observed transport at least in the steep SS range of our InSe-Si heterojunction triode devices is dominated by BTBT. However, given that the ON current is also reducing with temperature further theoretical and experimental studies are necessary to confirm the exact mechanism of transport away from the SS region.

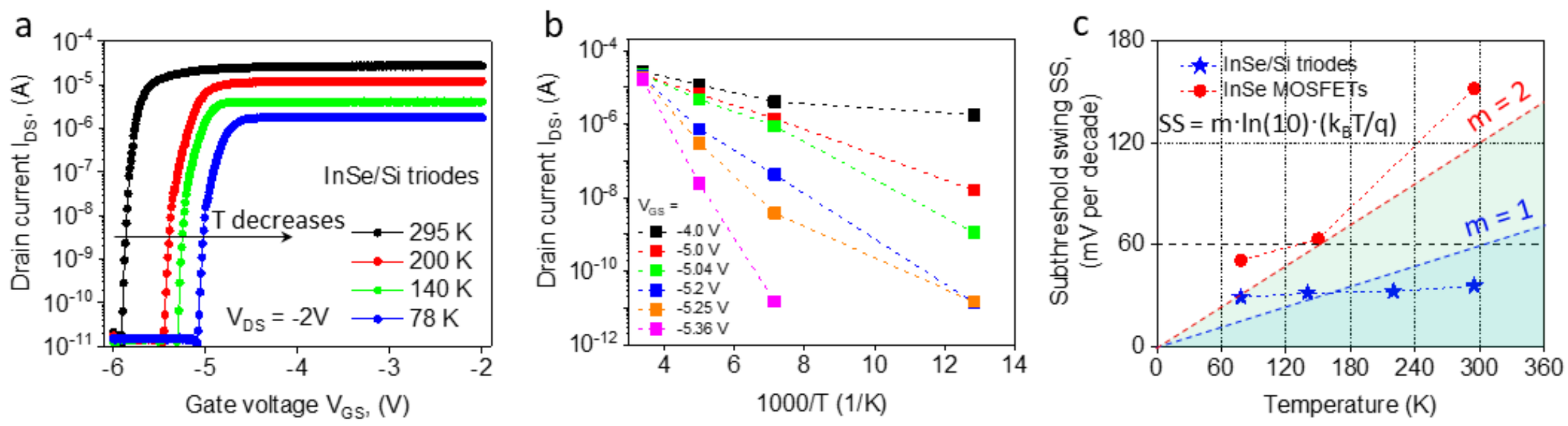

**Figure 4. Temperature-dependent electrical characteristics of InSe/Si 2D/3D triodes.** (a) Temperature-dependent $I_{DS}$-$V_{GS}$ transfer characteristics measured at $V_{DS}$ = -2.0 V. (b) Arrhenius plot of drain current as a function of 1000/T for various gate voltages. (c) Subthreshold swing as a function of temperature for InSe/Si 2D/3D heterojunction tunneling triodes and InSe control MOSFETs.

Finally, given the steep SS of our devices combined with the large ON/OFF ratio we perform benchmarking of our devices with literature precedent on select steep SS devices operating at room T. We compare our devices on three important metrics: (i) SS vs Drain current density, (ii) Drain current at SS = 60 mV/dec vs average SS and finally (iii) Current at SS= 60 mV/dec vs OFF current. Figure 5 shows values as a function of drain current for various steep SS TFET devices and negative capacitance FETs[23-26]. The SS versus $I_{DS}$ for our device is extracted from the transfer curves in Figure 2a. The comparison data clearly indicates that the InSe/Si TFETs simultaneously have small SS, high $I_{60}$ (current where SS becomes 60 mV/decade) and current density more than two orders of magnitude larger than that obtained in $MoS_2$/Ge TFET. $I_{60}$ is the chosen metric since the current saturates soon after that and would be point of operation of the ON state of the device. Further, we propose that this current density could be raised even further by removing the interfacial silicon oxide layer, as higher on-state current could improve the gate-to-source capacitance. Likewise, shrinking device dimensions and use of a graphene top contact can also help improve the $I_{60}$.

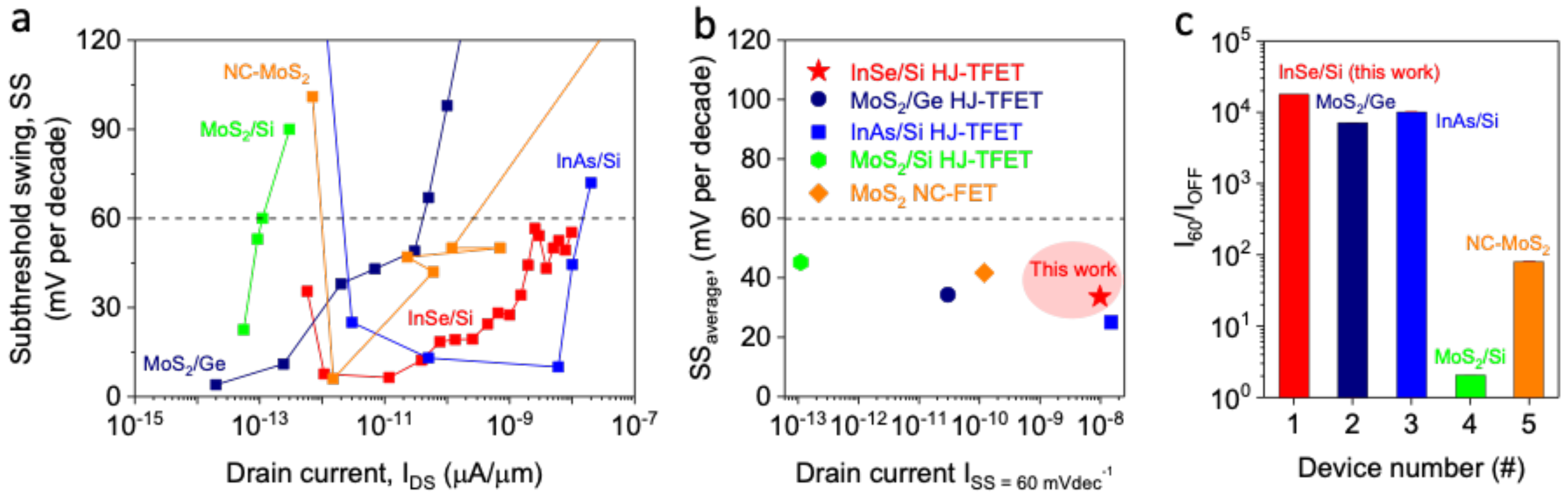

**Figure 5**. **Performance comparison of the InSe/Si 2D/3D HJ-TFETs with reported subthermionic HJ-TFETs and NC-FETs.** (a) Subthreshold swing as a function of drain current for various types of subthermionic FETs. The black dashed line represents the SS = 60 mV/decade limit. The red squares represent the SS values of InSe/Si HJ-TFETs in this work. The InSe/Si heterojunction devices show SS below 60 mV/decade for 4-5 orders of magnitude of drain current, with a minimum of around 6 mV/decade. (b) Average SS versus drain current where SS becomes lower than 60 mV/decade. Here, $SS_{average}$ of InSe/Si covers > 4 decades of $I_{DS}$ (red), $SS_{average}$ of $MoS_2$/Ge covers > 4 decades (navy), $SS_{average}$ of InAs/Si covers ~ 4 decades (blue), $SS_{average}$ of $MoS_2$/Si covers < 1 decade (green), $SS_{average}$ of NC-$MoS_2$ covers ~ 3 decades (orange). (c) $I_{60}/I_{OFF}$ for various subthermionic devices. The $I_{60}/I_{OFF}$ ratio of InSe/Si HJ-TFETs is > $10^4$.

In summary, we have demonstrated BTBT heterojunction field-effect transistors through vdW integration of 2D InSe on 3D silicon. Owing to the atomically thin nature of 2D InSe and fixed doping profile in 3D bulk silicon, bands can be effectively modulated by a capacitively coupled gate enabling strong modulation of band alignment resulting in direct BTBT. A minimum SS of 6.4 mV/dec with an average SS of 34 mV/decade over four decades of drain current at room T are reported in the as-fabricated InSe/Si 2D/3D heterojunction TFETs. In addition, the devices exhibit a high current on/off ratio of up to $10^6$ and on-state current density of 0.3 μA/μm at a $V_{DS}$ = -1 V. Our results suggest that 2D/3D integration is a viable path towards ultra-low-power and highly scaled logic switches. Further given that Indium has low solubility in Si[27] and InSe is lattice matched to Si [111][28,29], our approach presents a viable path toward large area growth and scaling of high performance InSe/Si heterojunction triodes for next generation digital logic.

## Methods:

**Device Fabrication**: All devices were fabricated using electron beam lithography (EBL) using a three-step process. Electron-beam resist (PMMA) was first spin-coated onto the $P^{++}$ Si substrate (WaferPro, B doping with resistivity of 0.005 ohm.cm) capped with a 50-nm $SiO_2$ layer grown using dry oxidation process. Then, a square pattern (size: 10 x 10 ums) was defined by EBL (step 1), following by a development process. After that, the $Si/SiO_2$ substrate was immersed into a buffered oxide etch (BOE) solution for about 1 min to etch the $SiO_2$ layer exposing the lower Si layer. Finally, the rest of the resist was removed by acetone, and the as-prepared substrate was immediately transferred to the glove box to avoid reoxidation of the exposed Si window. In the glove box, a few-layer 2D InSe flake was physically transferred onto the exposed Si window. The flake was exfoliated using a tape from the bulk crystal and then stuck onto a polydimethoxysilane (PDMS) stamp (GelPak, A4). The metal electrode patterns of the devices were then defined by EBL (step 2), following by metal deposition and lift-off processes. After that, we use the atomic layer deposition (ALD) method to deposit a 10 nm-thick $AlO_x$ top-gate dielectric on InSe. The top-gate electrode was finally defined by EBL (step 3), metal deposition and lift-off processes.

**Growth and characterization of InSe crystals**

InSe single crystals were grown by vertical Bridgman method using a non-stoichiometric polycrystalline $In_{1.04}Se_{0.96}$ charge. n- and p-type doping was achieved by adding Sn (2 at.%) and Zn (as ZnSe, 0.3 at.%), respectively, during InSe charge synthesis. The InSe melt was equilibrated at 720 °C for several hours, then the ampoule was translated across a temperature gradient at a rate of 0.5 mm/h.

**Electrical characterization:** The electrical measurements for all devices were performed by a Lakeshore probe station combined with a Keithley 4200 semiconductor analyzer. All the measurements were carried out at room temperature unless noted in the figures.

**Simulations:** TCAD Device simulations were performed using Synopsys Sentaurus Device Package. Additional details for the simulations are provided in Supporting information.

**Electron microscopy characterization:** Electron transparent cross-sectional sample was prepared with an FEI Nova NanoLab 600 DualBeam (SEM/FIB). A FEI Titan 80-300 probe-corrected STEM/TEM microscope operating at 300 keV was employed to acquire selected area electron diffraction patterns, STEM images and STEM EDS line-scan.

**Data availability:**

The data that support the conclusions of this study are available from the corresponding author upon reasonable request.

**Code availability:**

Any codes used in this study are available from the corresponding author upon request.

**Ethics declarations:**

**Competing interests**

D.J. and J.M. have filed an invention disclosure based on this work. The authors declare no other competing interests.

**Author contributions:**

D.J. and J.M. conceived the idea/concept. D.J. directed the collaboration and execution. J.M fabricated all devices with assistance from B.S. and measured them with assistance of X.L. Simulations were done by C.L. Electron microscopy and cross section cutting of the devices was performed by H.Z. InSe crystals were grown by S.K. A.V.D. supervised crystal synthesis, Hall measurements and electron microscopy. N.R.G. and T.B. performed photoemission spectroscopy measurements. J.M., C.L. and D.J. wrote the manuscript with contributions from all authors.

**Acknowledgements:**

D.J. and J.M. acknowledges primary support for this work by the Air Force Office of Scientific Research (AFOSR) FA9550-21-1-0035. D.J. also acknowledge partial support from Intel Rising Star Award and the University Research Foundation (URF) at Penn. D.J. also acknowledges partial support from U.S. Army Research Office under contract number W911NF-19-1-0109 and support from University of Pennsylvania Materials Research Science and Engineering Center (MRSEC) (DMR-1720530) in addition to usage of MRSEC supported facilities. The sample fabrication, assembly and characterization were carried out at the Singh Center for Nanotechnology at the University of Pennsylvania which is supported by the National Science Foundation (NSF) National Nanotechnology Coordinated Infrastructure Program grant NNCI-1542153. A.V.D. and S.K. acknowledge the support of Material Genome Initiative funding allocated to NIST. H.Z. acknowledges support from the U.S. Department of Commerce, National Institute of Standards and Technology under the financial assistance awards 70NANB19H138. T.B. and N.R.G. acknowledge support from the Air Force Office of Scientific Research under Award No. FA9550-19RYCOR050.

**Disclaimer:** Certain commercial equipment, instruments, or materials are identified in this paper in order to specify the experimental procedure adequately. Such identification is not intended to imply recommendation or endorsement by the National Institute of Standards and Technology, nor is it intended to imply that the materials or equipment identified are necessarily the best available for the purpose.

**References:**

1 Iannaccone, G., Bonaccorso, F., Colombo, L. & Fiori, G. Quantum engineering of transistors based on 2D materials heterostructures. *Nature Nanotechnology* **13**, 183-191, doi:https://doi.org/10.1038/s41565-018-0082-6 (2018).

2 Li, X. *et al.* in *Beyond-CMOS Technologies for Next Generation Computer Design* (eds Rasit O. Topaloglu & H. S. Philip Wong) 195-230 (Springer International Publishing, 2019).

3 Lu, H. & Seabaugh, A. Tunnel Field-Effect Transistors: State-of-the-Art. *IEEE Journal of the Electron Devices Society* **2**, 44-49, doi:10.1109/JEDS.2014.2326622 (2014).

4 Fiori, G. *et al.* Electronics based on two-dimensional materials. *Nature Nanotechnology* **9**, 768-779, doi:10.1038/nnano.2014.207 (2014).

5 Yan, X. *et al.* Tunable SnSe2/WSe2 Heterostructure Tunneling Field Effect Transistor. *Small* **13**, doi:10.1002/smll.201701478 (2017).

6 T, R. *et al.* Dual-gated MoS2/WSe2 van der Waals tunnel diodes and transistors. *ACS nano* **9**, doi:10.1021/nn507278b (2015).

7 Britnell, L. *et al.* Field-Effect Tunneling Transistor Based on Vertical Graphene Heterostructures. *Sciene* **335**, doi:10.1126/science.1218461 (2012).

8 Nourbakhsh, A., Zubair, A., Dresselhaus, M. S. & Palacios, T. Transport Properties of a MoS2/WSe2 Heterojunction Transistor and Its Potential for Application. *Nano Letters* **16**, 1359–1366, doi:10.1021/acs.nanolett.5b04791 (2016).

9 Kim, S. *et al.* Thickness-controlled black phosphorus tunnel field-effect transistor for low-power switches | Nature Nanotechnology. *Nature Nanotechnology* **15**, 203-206, doi:10.1038/s41565-019-0623-7 (2020).

10 Vandenberghe, W. G. *et al.* Figure of merit for and identification of sub-60 mV/decade devices. *Applied Physics Letters* **102**, doi:10.1063/1.4773521 (2013).

11 Lv, R. *et al.* Two-dimensional transition metal dichalcogenides: Clusters, ribbons, sheets and more. *Nano Today* **10**, 559-592 (2015).

12 K.S., N., A., M., A., C. & A.H., C. N. 2D materials and van der Waals heterostructures. *Science* **353**, doi:10.1126/science.aac9439 (2016).

13 Bae, S.-H. *et al.* Integration of bulk materials with two-dimensional materials for physical coupling and applications. *Nature Materials* **18**, 550-560, doi:https://doi.org/10.1038/s41563-019-0335-2 (2019).

14 Akinwande, D. *et al.* Graphene and two-dimensional materials for silicon technology. *Nature* **573**, 507-518, doi:https://doi.org/10.1038/s41586-019-1573-9 (2019).

15 D, J., T.J., M. & M.C., H. Mixed-dimensional van der Waals heterostructures. *Nature materials* **16**, 170-181, doi:10.1038/nmat4703 (2017).

16 Miao, J. *et al.* Gate-Tunable Semiconductor Heterojunctions from 2D/3D van der Waals Interfaces. *Nano Letters* **20**, 2907-2915, doi:10.1021/acs.nanolett.0c00741 (2020).

17 Murali, K., Dandu, M., Das, S. & Majumdar, K. Gate-Tunable WSe2/SnSe2 Backward Diode with Ultrahigh-Reverse Rectification Ratio. *Applied Material Interfaces* **10**, 5657-5664, doi:10.1021/acsami.7b18242 (2018).

18 Wang, C. *et al.* Gate-tunable diode-like current rectification and ambipolar transport in multilayer van der Waals ReSe2/WS2 p–n heterojunctions. *Physical Chemistry Chemical Physics* **18**, 27750-27753, doi:10.1039/C6CP04752A (2016).

19 Sangwan, V. K. & Hersam, M. C. Electronic Transport in Two-Dimensional Materials. *Annual Review of Physical Chemistry* **69**, 299-325, doi:10.1146/annurev-physchem-050317-021353 (2018).

20 Jariwala, D. *et al.* Band-like transport in high mobility unencapsulated single-layer MoS2 transistors. *Applied Physics Letters* **102**, doi:10.1063/1.4803920 (2013).

21 Martinez-Pastor, J., Segura, A., Julien, C. & Chevy, A. Shallow-donor impurities in indium selenide investigated by means of far-infrared spectroscopy. *Physical Review B* **46**, 4607 (1992).

22 Hopkinson, D. G. *et al.* Formation and healing of defects in atomically thin GaSe and InSe. *ACS nano* **13**, 5112-5123 (2019).

23 Sarkar, D. *et al.* A subthermionic tunnel field-effect transistor with an atomically thin channel. *Nature* **526**, 91-95, doi:doi:10.1038/nature15387 (2015).

24 Si, M. *et al.* Steep-slope hysteresis-free negative capacitance MoS2 transistors. *Nature Nanotechnology* **13**, 24-28, doi:doi:10.1038/s41565-017-0010-1 (2018).

25 Tomioka, K., Yoshimura, M. & Fukui, T. in *Symposium on VLSI Technology (VLSIT)* 47-48 (2012).

26 Hyuk Shin, G. *et al.* Vertical-Tunnel Field-Effect Transistor Based on a Silicon–MoS2 Three-Dimensional–Two-Dimensional Heterostructure. **10**, 40212–40218, doi:10.1021/acsami.8b11396 (2018).

27 Keck, P. H. & Broder, J. The Solubility of Silicon and Germanium in Gallium and Indium. *Physical Review* **90**, doi:10.1103/PhysRev.90.521 (1953).

28 Higashi, G. S., Becker, R. S., Chabal, Y. J. & Becker, A. J. Comparison of Si(111) surfaces prepared using aqueous solutions of NH4F versus HF. *Applied Physics Letters* **58**, 1656, doi:10.1063/1.105155 (1991).

29 O. Madelung, U. R., M. Schulz. in *Non-Tetrahedrally Bonded Elements and Binary Compounds I* Vol. 41C (SpringerMaterials, 1998).

**Supplementary Information**

# 2D Metal Selenide-Silicon Steep Sub-Threshold Heterojunction Triodes with High On-Current Density

Jinshui Miao,[1,] Chloe Leblanc,[1] Xiwen Liu,[1] Baokun Song,[1] Huairuo Zhang,[2, 3] Sergiy Krylyuk,[2] Albert V.Davydov,[2] Tyson Back,[4] Nicholas Glavin,[4] Deep Jariwala[1,]*

[1] Department of Electrical and Systems Engineering, University of Pennsylvania, Philadelphia PA 19104, United States

[2] Materials Science and Engineering Division, National Institute of Standards and Technology, Gaithersburg, MD 20899, United States

[3] Theiss Research, Inc., La Jolla, CA 92037, United States

[4] Air Force Research Laboratory, Materials and Manufacturing Directorate, Wright-Patterson AFB, OH 45433

* Corresponding author: dmj@seas.upenn.edu

**Section S1: Hall effect measurements, photoemission and transmission electron microspectroscopy**

**1. Hall Measurements on InSe bulk crystals for various doping conditions.**

Samples for Hall effect measurements were cleaved with a razor blade and cut into parallelepipeds with about 0.1 mm in thickness and 8 mm x 8 mm in size. Ohmic contacts to InSe were made by soldering high-purity In. Hall effect measurements were carried out in the Van der Pauw geometry at room temperature.

**Table 1. Summary of Hall effect measurements on InSe crystals.**

| | **Undoped InSe** | **n-doped InSe** | **p-doped InSe** |
|---|---|---|---|
| **Dopant, Concentration in the melt** | n/a | Sn, 2.6 at.% | Zn (from ZnSe), 0.3 at.% |
| **Carrier bulk concentration, $cm^{-3}$** | $3.16 \cdot 10^{14}$ | $1.47 \cdot 10^{16}$ | $7.9 \cdot 10^{13}$ |
| **Hall Mobility, $cm^2/(V \cdot s)$** | 1390 | 616 | 43 |

**2. Photoemission measurements and X-TEM composition profile.**

X-ray photoelectron spectroscopy was carried out on a Kratos AXIS Ultra using a monochromatic Al Kα (1486.6 eV) X-ray source. The voltage and emission current on the X-ray source was 12 keV and 10 mA respectively. Scans to determine the VBM were acquired taking 0.1 eV steps with 100 ms dwell time and a 10 eV pass energy. The VBM was determined by a linear extrapolation to the binding energy axis.

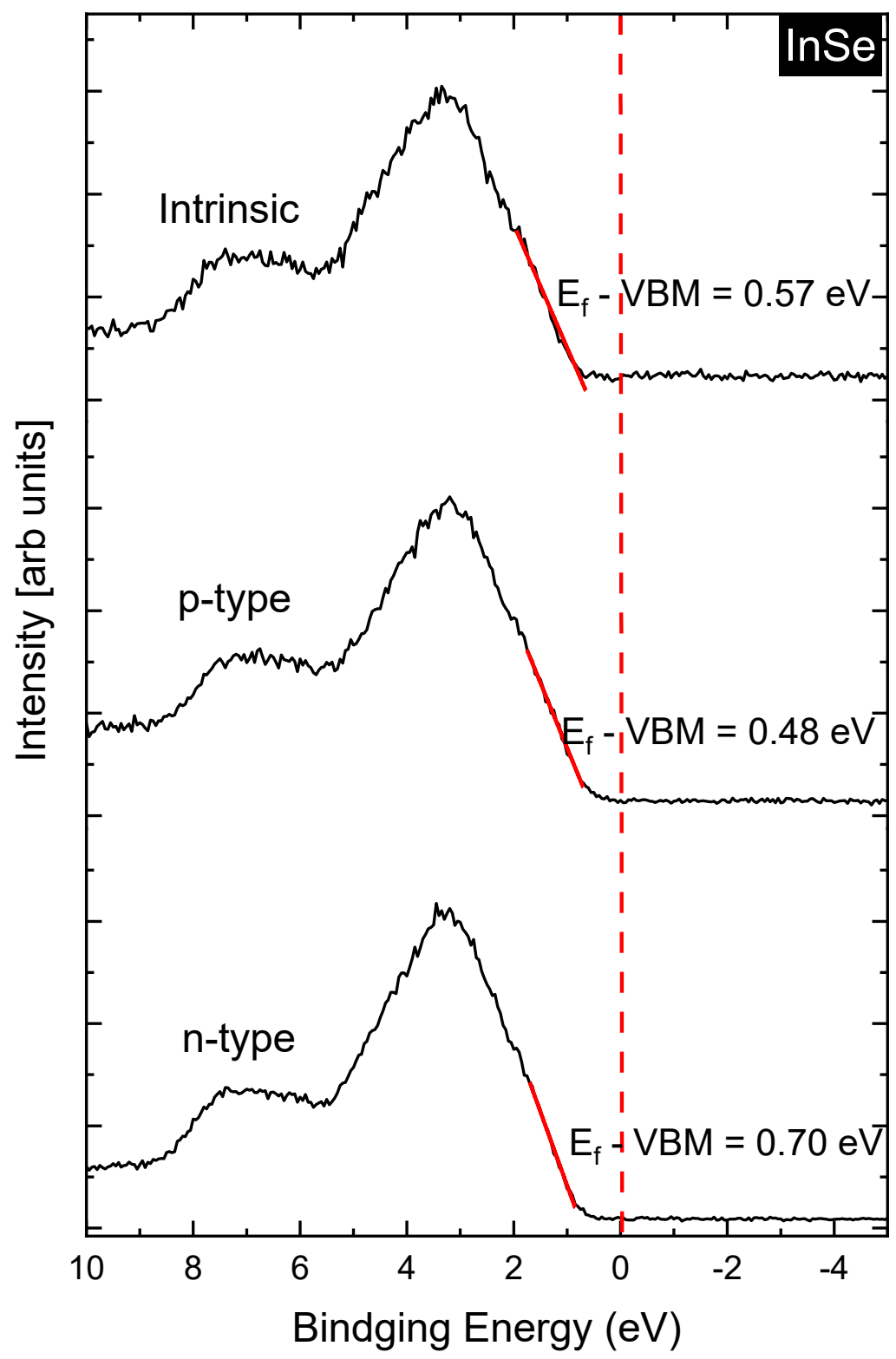

Figure S1-1: Photoemission spectra from bulk InSe crystals with various doping conditions. The $E_f$-VBM energies are shown.

**3. Details on Cross section cutting and milling as well as compositional and crystal structure analysis.**

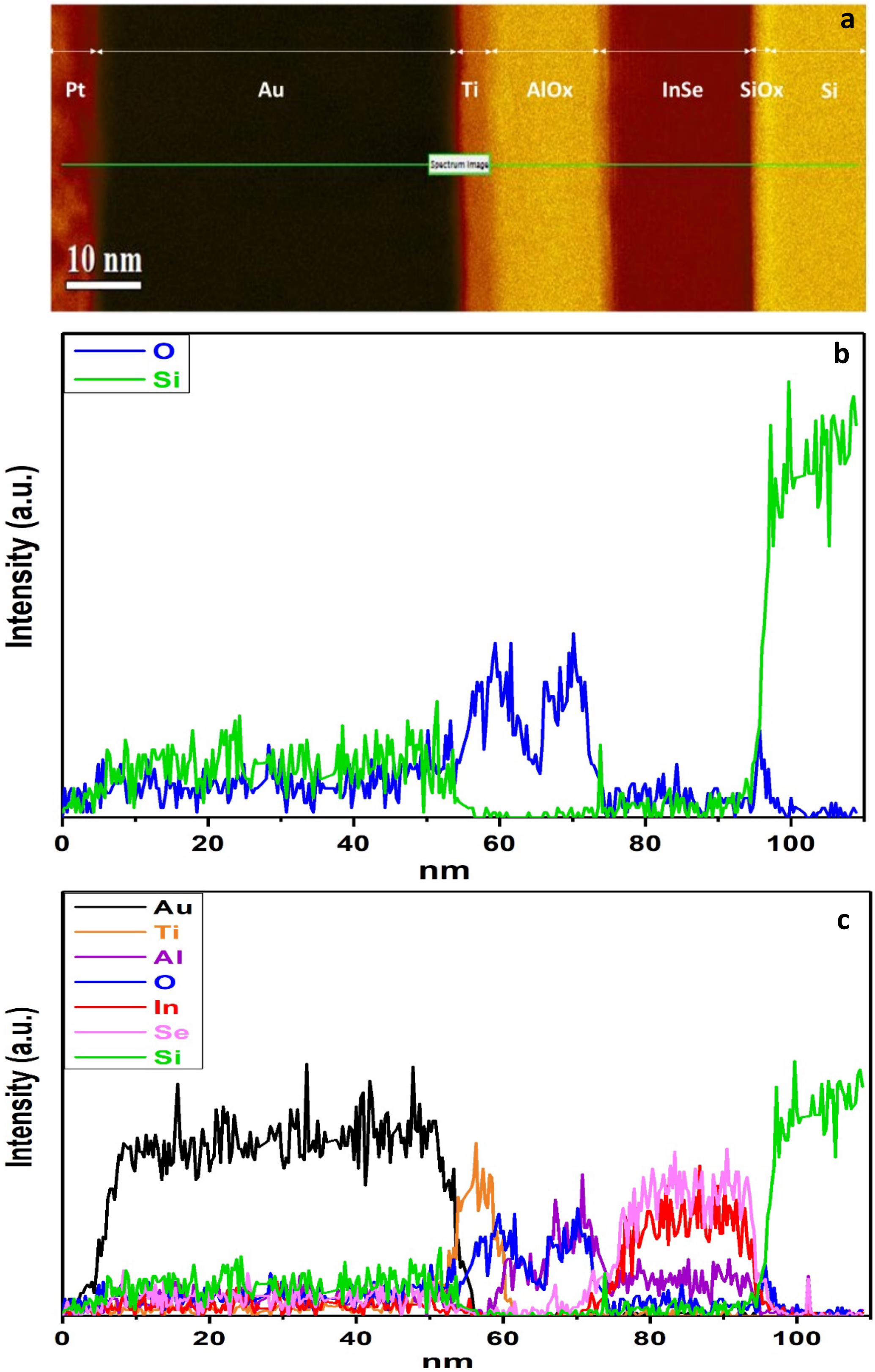

Figure S1-2. Cross sectional composition analysis of the InSe/Si heterojunction triode device. a. Bright field cross section micrograph of the device showing various layers in the heterojunction and the gate stack. A ~2 nm thick SiOx layer is identified at the interface between InSe and Si. b. Line cuts of composition as a function of depth of the device from the top for oxygen (blue) and silicon (green). The clear peak in oxygen at the start of Si edge shows the presence of a thin sub-oxide layer. c. Composition line profiles for all relevant elements indicating the appropriate layer thicknesses and compositions.

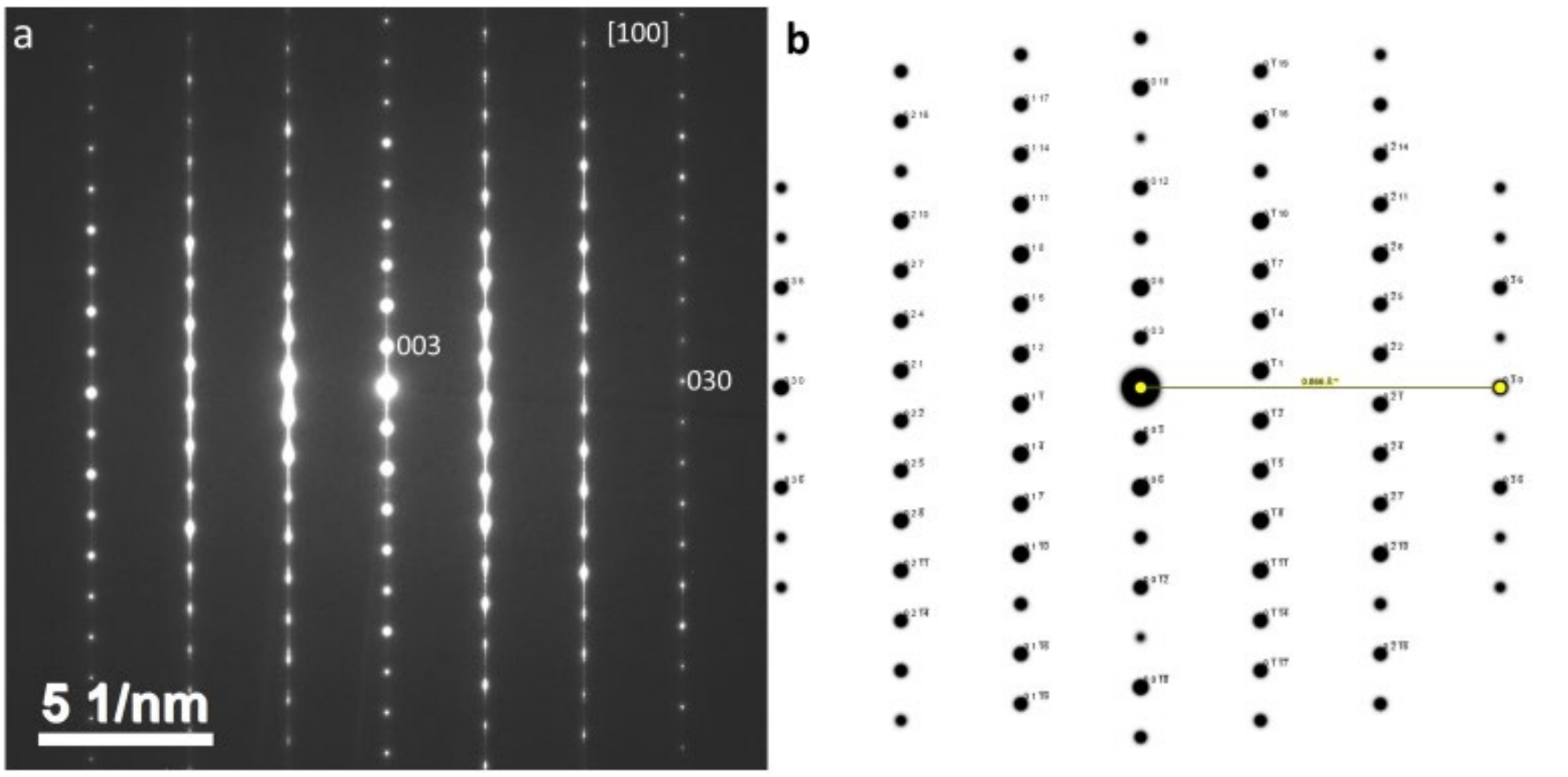

Figure S1-3. a. Experimentally measured electron diffraction pattern of gamma (γ) phase InSe along [100] direction. b. Simulated electron diffraction pattern that matches the experiment. These results suggest that the crystal flake used in the devices is definitely gamma (γ) phase InSe.

## Section S2. TCAD simulation parameters and methods:

Synopsys Sentaurus Device was used for the TCAD simulation. Sentaurus Device predicts the behavior of devices by discretizing the device architecture and information to solve it in partial differential equation form. Our model starts with the 3D definition of the fabricated device geometry along with the location of the contacts and their applied voltage. A separate Physics section records the phenomena we wish to take into consideration, such as recombination and band-to-band tunneling. Finally, we instruct the model how to measure its simulated device—which predefined electrode to sweep over, over what range, using which equations, etc. The model can also plot a wide array of elements such as bandgaps, electron and hole distribution and current, electric potential among other things, all of which along the 3D direction of choice. Our model solves sets of coupled differential equations namely Poisson’s equation, continuity equations for electrons and holes and current equations for electrons and holes in two dimensions.

The materials parameters used for the simulations are detailed in the table below.

**Table 2. Junction Materials parameters for TCAD simulations**

| Parameters | InSe | Si | $SiO_2$ |
|---|---|---|---|
| Electron affinity | 4.6 eV[1] | 4.0727 eV[2] | 0.4 eV[3] |
| Bandgap | 1.27 eV[4] | 1.12416[2] | 9[5] |
| Permittivity | $7\varepsilon_0$[6] | $11.7\varepsilon_0$[7] | $3.9\varepsilon_0$[5] |
| Anisotropic permittivity | *undefined* | $11.7\varepsilon_0$[7] | *undefined* |
| Lattice thermal conductivity | 0.259 W/(K cm)[8] | 1.6964 W/(K cm)[9] | 0.014 W/(K cm)[10] |
| Lattice heat capacity | 1.479 J/(K cm^3)[11] | 1.63 J/(K cm^3)[7] | 1.67 J/(K cm^3) |
| Effective mass (DOS) _ electron | 0.18[12] | 0.36[13] | 0.3[14] |
| Effective mass (DOS)_ hole | 0.5[12] | 0.81[13] | 0.33[15] |

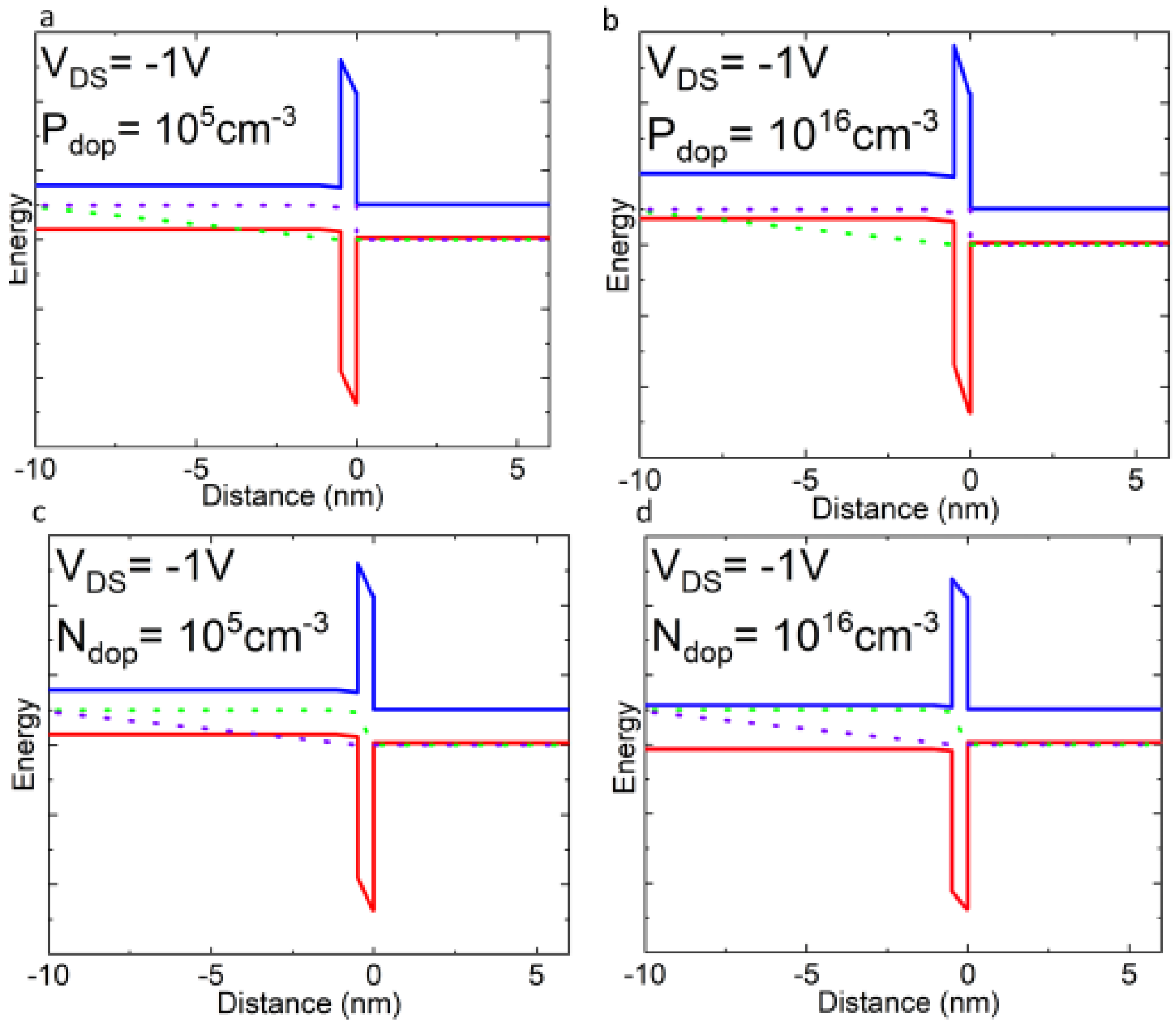

**Figure S2-1.** Band diagram representation of the InSe/Si 2D/3D tunneling FET at room temperature and various doping levels and species. The degenerately doped Si acts as the source and the InSe is the drain. (a) Lightly P-doped. (b) Heavily P-doped. (c) Lightly N-doped. (d) Heavily N-doped.

**Figure S2-1** represents the drain-to-source band-diagram of the InSe/Si 2D/3D heterojunction FET at different drain doping levels. For source region, degenerate p-type doping pins the Fermi level to the valence band. For the drain region, increasing boron doping raises the bands and causes the Fermi level to approach the conduction band, shown in **Figure S2-1 a and b**. The inverse effect happens with arsenic doping (**Figure S2-1 c and d**), which decreases the drain band level and causes the Fermi level to approach the valence band. This lowering facilitates the tunneling of electrons from source to drain, as represented by the black arrow in Figure S1d. The heavier the doping, the easier it is for the electrons to overcome the oxide barrier and tunneling to occur.

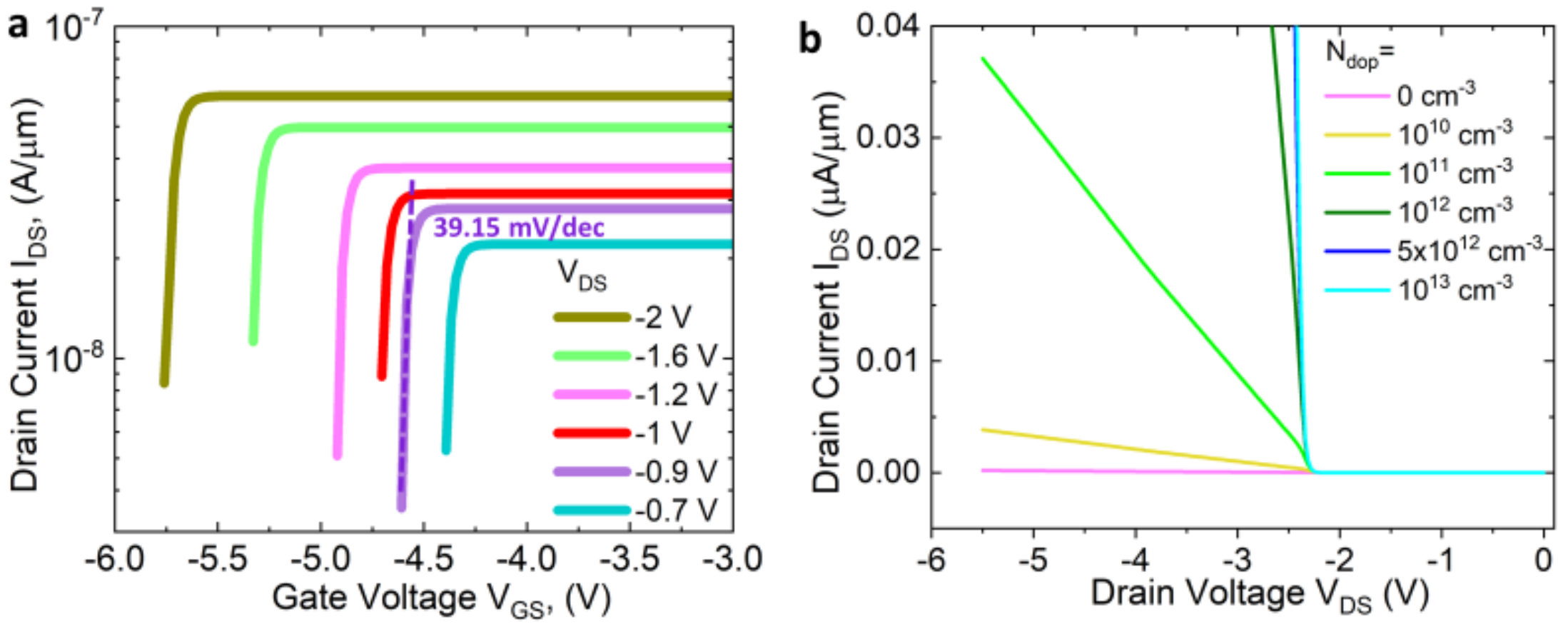

**Figure S2-2**. a. Simulated transfer characteristis of the InSe-Si heterojunction triodes at various drain voltages showing SS < 60 mV/dec. b. Simulation of the linear-scale $I_{DS}$-$V_{DS}$ output characteristics of InSe/Si heterojunction device at room temperature and various $V_{GS}$ values. The $SiO_2$ layer thickness set to 2 nm in this simulation.

The simulations in **Figure S2-2** accurately predict the qualitative behavior observed in Figure 2d and Figure 3d respectively.

## Section S3. Electrical Characteristics of $WSe_2$/Si $P^{++}$ heterojunctions.

Devices were also made from $WSe_2$ and Si $P^{++}$ wafers using exactly the same procedure as outlined for InSe based devices. The bulk crystals of $WSe_2$ were purchased from HQ graphene. Control $WSe_2$ FETs were also fabricated from the same flakes which show SS values > 100 mV/dec. across all range of currents.

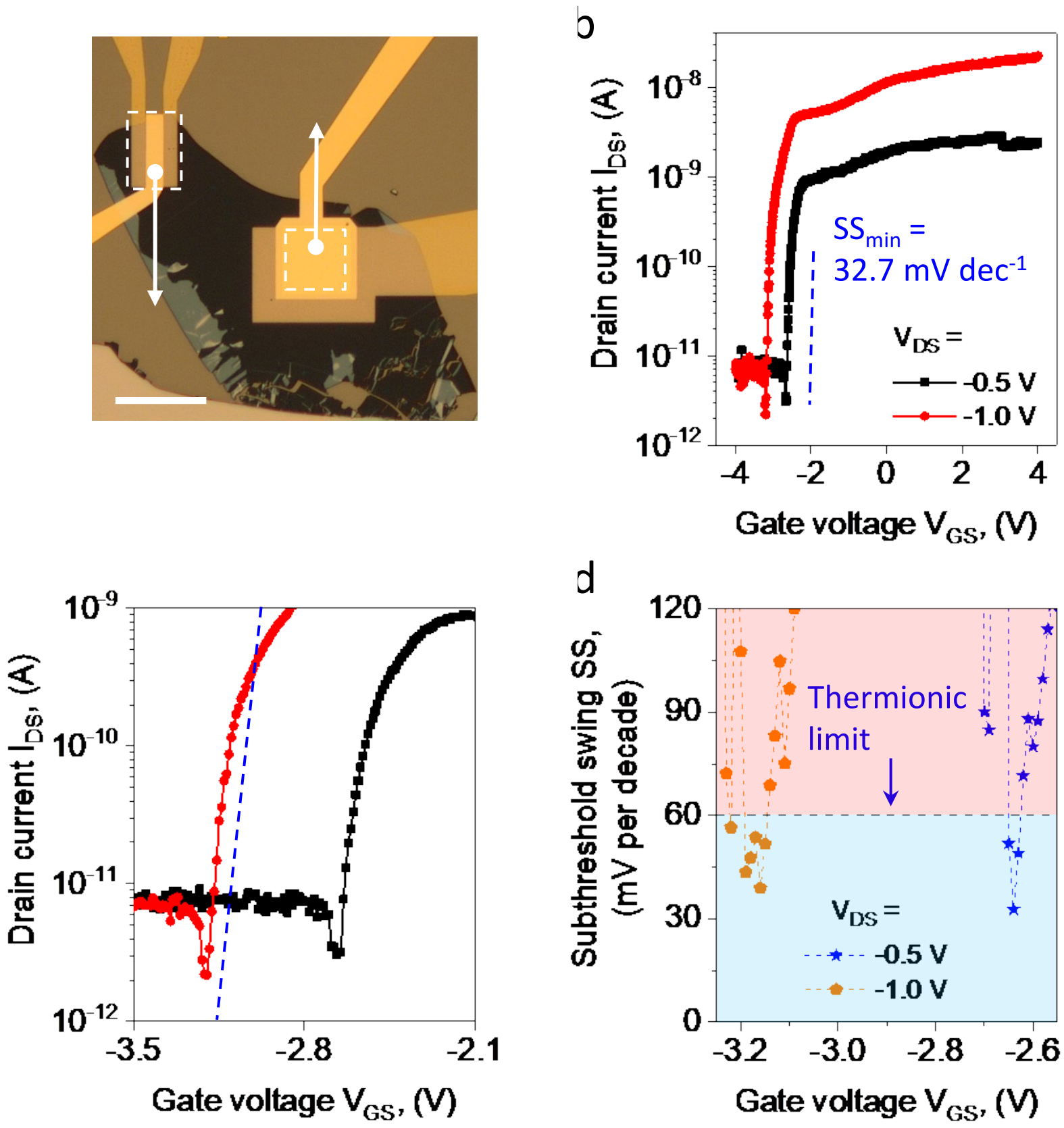

**Figure S3**. $WSe_2/P^{++}$ Si heterojunction tunnel field-effect transistors (T-FET) and their electrical characterizations. (a) Optical micrograph of a representative $WSe_2/p^{++}$ Si 2D/3D heterojunction T-FET device. Scale bar: 10 um. (b) Logarithmic-scale $I_{DS}$-$V_{GS}$ transfer characteristics of $WSe_2/p^{++}$ Si heterojunction TFET measure at $V_{DS}$=-0.5 V. (c) Zoom in $I_{DS}$-$V_{GS}$ transfer characteristics of $WSe_2/p^{++}$ Si heterojunction devices measured at room temperature with applied $V_{DS}$ = -0.5 V (black) and -1.0 V (red) plots. The minimum sub-threshold swing (SS) value is 32.7 mV/dec. (d) Subthreshold swing as a function of gate voltage ($V_{GS}$) with applied $V_{DS}$ = -0.5 V (blue) and -1.0 V (orange) plots. The black dashed line shows the thermionic limit of 60 mV/dec.

**Section S4. Control InSe MOSFETs and their temperature dependent transport properties.**

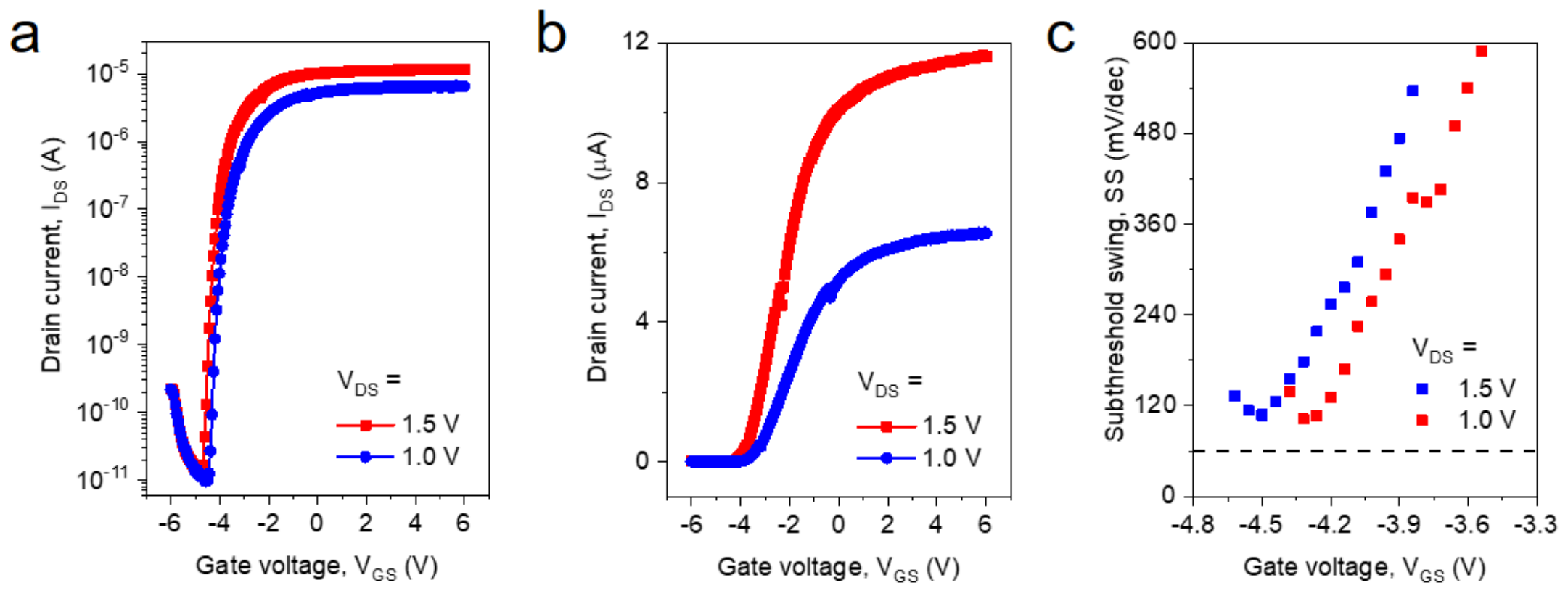

**Figure S4-1**. InSe field-effect transistor characterizations at room temperature. (a) Logarithmic-scale and (b) linear-scale $I_{DS}$-$V_{GS}$ transfer characteristics of InSe FET measure at $V_{DS}$=1.5 V (red) and 1.0 V (blue). (c) Subthreshold swing as a function of gate voltage ($V_{GS}$) with different applied $V_{DS}$ plots. The black dashed line shows the thermionic limit of 60 mV/dec at room temperature.

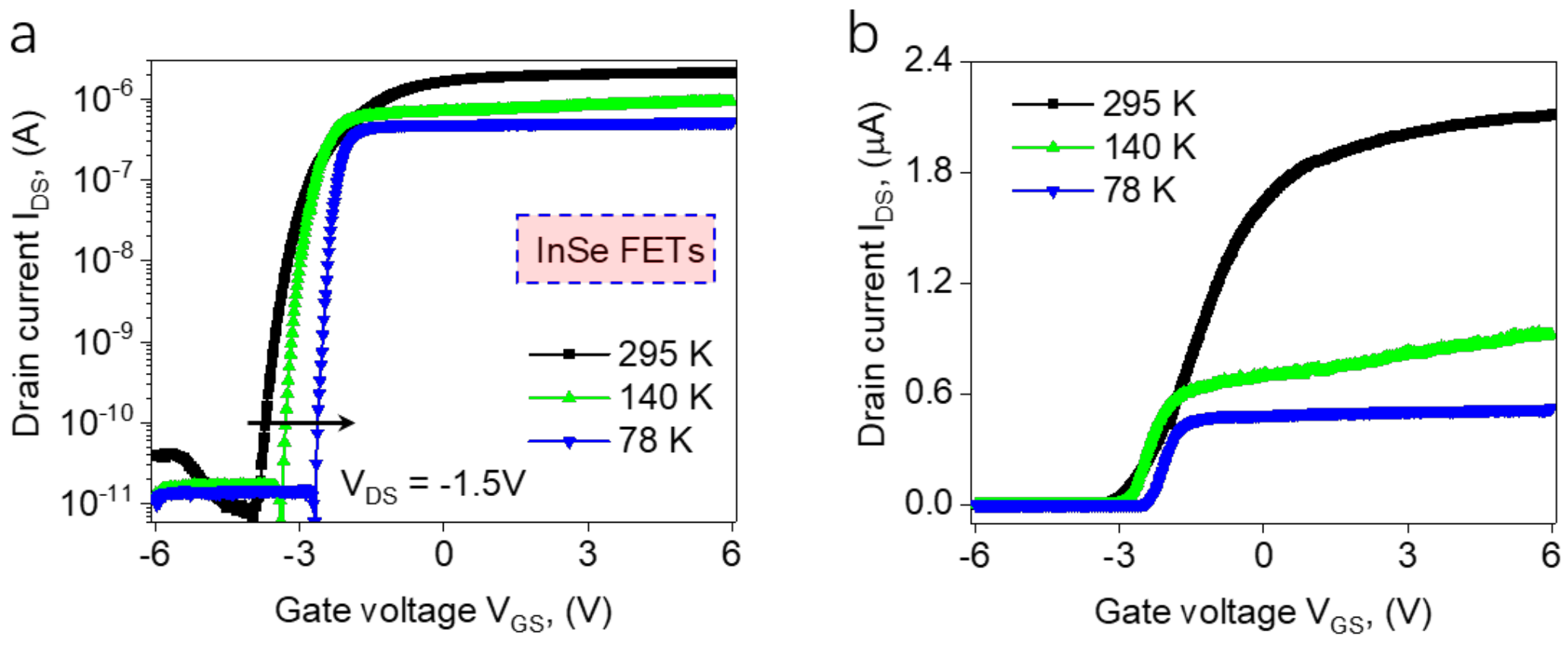

**Figure S4-2**. InSe field-effect transistors characterizations. (a) Temperature-dependent logarithmic-scale and (b) linear-scale $I_{DS}$-$V_{GS}$ transfer characteristics of InSe FET measure at $V_{DS}$=-1.5 V. A clear decrease in both SS and on current is observed. The decrease in SS suggests thermionic transport while the decrease in ON current suggests imperfect contacts with a thermal barrier between the metal and the semiconductor.

## Section S5: Additional InSe/Si $P^{++}$ heterojunction transfer characteristics.

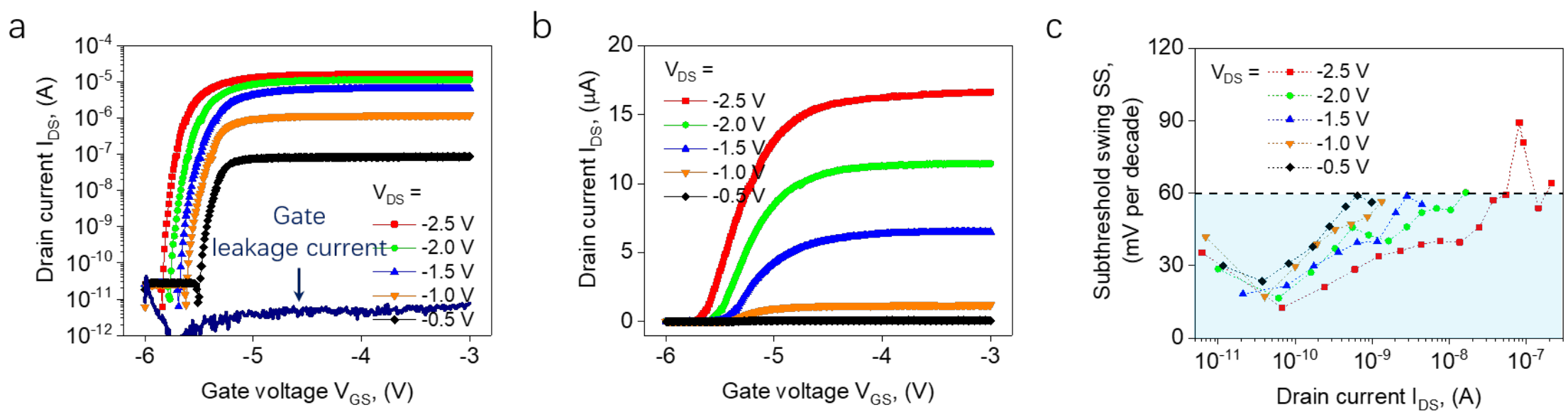

**Figure S5**. Room-temperature electrical characteristics of InSe/Si heterojunction field-effect transistors (TFET). (a) Logarithmic-scale $I_{DS}$-$V_{GS}$ transfer characteristics of InSe/Si heterojunction TFET measured at different $V_{DS}$. The blue line represents gate leakage current of around $10^{-12}$ A. (b) Linear-scale $I_{DS}$-$V_{GS}$ transfer characteristics measured at different $V_{DS}$. (c) Subthreshold swing as a function of drain current ($I_{DS}$) with different applied $V_{DS}$ plots. The black dashed line shows the thermionic limit of 60 mV/dec.

## References:

1. Mudd, G. W.; Svatek, S. A.; Hague, L.; Makarovsky, O.; Kudrynskyi, Z. R.; Mellor, C. J.; Beton, P. H.; Eaves, L.; Novoselov, K. S.; Kovalyuk, Z. D.; Vdovin, E. E.; Marsden, A. J.; Wilson N. R.; Patanè, A., High Broad-Band Photoresponsivity of Mechanically Formed InSe–Graphene van der Waals Heterostructures. *Advanced Materials*: 2015; Vol. 27, pp 3760-3766.

2. Braunstein, Moore, and Herman in C. Penn, T. Frommherz, and G. Bauer, Properties of Silicon Germanium and SiGe: Carbon EMIS Datareviews Series, no. 24, chapter 4.1

3. Tashmukhamedova, D.A., Yusupjanova, M.B. Emission and Optical Properties of SiO2/Si Thin Films. *Journal of Surface Investigation: X-ray, Synchrotron and Neutron Techniques*: 2016; Vol. 10, No. 6, pp. 1273-1275.

4. Mudd, G. W.; Svatek, S. A.; Ren, T.; Patanè, A.; Makarovsky, O.; Eaves, L.; Beton, P. H.; Kovalyuk, Z. D.; Lashkarev, G. V.; Kudrynskyi, Z. R.; Dmitriev, A. I., Tuning the Bandgap of Exfoliated InSe Nanosheets by Quantum Confinement. *Advanced Materials*: 2013; Vol. 25, pp 5714–5718.

5. G. Neudeck. *The PN Junction Diode* (Addison-Wesley Publishing Co., Reading, Massachusetts, 1989)

6. Politano, A.; Campi, D.; Cattelan, M.; Ben Amara, I.; Jaziri, S.; Mazzotti, A.; Barinov, A.; Gürbulak, B.; Duman, S.; Agnoli, S.; Caputi, L. S.; Granozzi, G.; Cupolillo, A. Indium selenide: an insight into electronic band structure and surface excitations *Scientific Reports* 2017, **7.**

7. Schaffler, Properties of Advanced Semiconductor Materials GaN, AlN, InN, BN, SiC, SiGe, John Wiley & Sons, Inc., New York, pp. 149-188, 2001

8. Wang, Q.; Han, L.; Wu, L.; Zhang, T.; Li, S.; Lu, P. Strain Effect on Thermoelectric Performance of InSe Monolayer. *Nanoscale Research Letters* **2019,** 14.

9. Glasbrenner and Slack, Phys. Rev., 164, 4A, pp 1058-1069, 1964

10. Zhu, W.; Zheng, G.; Cao, S.; He, H. Thermal conductivity of amorphous SiO2 thin film: A molecular dynamics study. *Scientific Reports* **2018,** 8.

11. Tyurin, A. V.; Gavrichev, K. S.; Zlomanov, V. P. Low-Temperature Heat Capacity and Thermodynamic Properties of InSe. *Inorganic Materials* **2007,** 43, (9), 921-925.

12. Segura, A. Layered Indium Selenide under High Pressure: A Review. *Crystals* **2018,** 8, (5).

13. Silicon - Band structure and carrier concentration. ioffe.ru.

14. König D., Rennau, M., Henker, M. Direct tunneling effective mass of electrons determined by intrinsic charge-up process. *Solid State Electronics* **2007**, 51, pp 650-654.

15. Vexler, M. I.; Tyaginov, S.; Shulekin, A. F. Determination of the hole effective mass in thin silicon dioxide film by means of an analysis of characteristics of a MOS tunnel emitter transistor. *Journal of Physics Condensed Matter* **2005,** 17, (50).